\documentclass[prl,aps,twocolumn,draft,showpacs]{revtex4}

\usepackage{graphicx}

\begin{document}

\title{Interplay of Rayleigh and Peierls Instabilities in Metallic Nanowires}

\author{D.~F.~Urban and Hermann~Grabert}

\affiliation{Physikalisches Institut,
Albert-Ludwigs-Universit\"at, D-79104 Freiburg, Germany}

\date{\today}

\begin{abstract}
A quantum-mechanical stability analysis of metallic nanowires
within the free-electron model is presented. The stability is
determined by an interplay of electron-shell effects, the
Rayleigh instability due to surface tension, and the Peierls
instability. Although the latter effect limits the maximum length
also for wires with ``magic radii", it is found that nanowires in
the micrometer range can be stable at room temperature.
\end{abstract}


\maketitle
\vskip2pc

Metal nanowires represent the ultimate limit of conductors down
to a single atom in thickness. In recent years, experimental
research on metal nanowires has burgeoned and  various techniques
have been employed to produce them for many different metals
\cite{Agr93,Yan99a,Yan00,Kon97,Kon00}. Remarkably, these
nanowires are quite stable, although one might expect them to
break up because of surface tension. Such an instability arises
from classical continuum mechanics and is know as the Rayleigh
instability \cite{Cha81}.
Experimentally the stability was found to depend sensitively on
the geometry of the wire. Yanson \emph{et al.}
\cite{Yan99a,Yan00} have measured conductance histograms for
alkali metal nanowires. At low temperature only wires with
certain ``magic radii" were stable.
In another notable experiment \cite{Kon97,Kon00}, Kondo \emph{et
al.} have produced long gold nanowires that were almost perfectly
cylindrical in shape. Electron microscope images show lengths
between 3 and 15 nm and diameters from 0.6 to 1.5 nm. A histogram
was built which again reflects stability for certain diameters
only.

Many features of metallic nanowires can be explained in terms of
a model of free electrons confined to the geometry of the wire
\cite{Footnote1}. At least for alkali metals, experiments on
conductance quantization and current noise are in good agreement
with theoretical predictions based on the Jellium model
\cite{Tor93,Sta97a,Kas99,Bue99a,Bue99b}. Furthermore, Kassubek
\emph{et al.} \cite{Kas01} have shown that a semi-classical
calculation of the density of states of nanowires leads to stable
magic radii, like those observed in experiments
\cite{Yan99a,Yan00}. However, a puzzling conclusion of the
semiclassical theory is that nanowires with a magic radius remain
stable whatever their length.

There is substantial interest in the stability of nanowires,
since they may serve as conductors in future nanocircuits. In
particular, one would like to know whether they remain stable up
to a length useful for applications. We therefore reexamine this
problem using a fully quantum-mechanical analysis.  It turns
out that the suppression of the Rayleigh instability by electron
shell effects, which is captured by the semiclassical theory, is
now supplemented by an interplay between the Rayleigh and a novel
Peierls-type instability missed in previous work. In fact, this
latter quantum mechanical instability limits the maximal length
of stable nanowires but still allows for wires in the micrometer
range sufficient for most applications.

%====================================================================
%
%                    M E T H O D S
%
%====================================================================

We treat the  metallic nanowire as an open system connected to
macroscopic metallic electrodes at each end. It is most naturally
described within a scattering matrix approach \cite{Sta97a}. The
relevant thermodynamic potential is the grand canonical potential
$\Omega$ which is related to the density of states $D(E)$  by
\begin{equation}
\label{gl::OmegaVonD}
    \Omega=-k_{B} T \int \!dE\; D(E) \;
    \ln\!\left[1+e^{-\frac{(E-\mu)}{k_{\mbox{\tiny B}}\,T}}
        \right],
\end{equation}
where $k_B$ is the Boltzmann constant, $T$ is the temperature and
$\mu$ is the chemical potential specified by the macroscopic
electrodes. The density of states can be calculated from the
scattering matrix $S(E)$ by
\begin{eqnarray}
\label{gl::DausS}
        D(E) &=&
    \frac{1}{2\pi i}\;\mbox{Tr}\left\{
    S^{\dagger}(E)\frac{\partial S}{\partial E} -
        \frac{\partial S^{\dagger}}{\partial E}S(E)\right\}.
\end{eqnarray}
A factor of 2 for spin degeneracy has been included. In order to
examine the stability of a nanowire, we determine the energy
change $\delta\Omega$ caused by a small deformation of the wire
geometry. A nanowire is only stable if all small deformations
result in an increase of $\Omega$.

%===========================================
%
%  section{Coupling}
%
%===========================================

The Jellium model considers free electrons that are kept in the
wire by a confining potential chosen as a hard-wall boundary. The
corresponding Schr\"odinger equation can be solved by a
separation ansatz
\begin{eqnarray}
\label{gl::Seperationsansatz}
        \Psi(r,\varphi,z)&=&\sum_\mu C_\mu(z)\chi_\mu(r,\varphi;z),
\end{eqnarray}
in cylindrical coordinates $r$, $\varphi$ and $z$. This is
motivated by the fact, that at a given position $z$ along
the wire we can find a complete set of orthogonal transverse
eigenfunctions $\chi_\mu(r,\varphi;z)$ with eigenenergies $E_\mu(z)$.

Inserting this ansatz into the Schr\"odinger
equation, we obtain  a system of coupled differential equations for
the coefficient functions $C_\mu(z)$ \cite{Ulr98}
\begin{eqnarray}
\label{gl::gekoppeltesDGLsys}
    &&\!\!\!\!\!\!\!\!\!\!\!
         C_\mu''(z)+ [E-E_\mu(z)] C_\mu(z)\;=\;
    \nonumber\\
    &&\!\!\!\!\!\!\!\!\!\!\!
    -\sum_\nu\left\{2A_{\mu\nu}(z)C_\nu'(z) +
        \left[ A_{\mu\nu}'(z) +
        A^2_{\mu\nu}(z)\right]C_\nu(z)\right\},\quad
\end{eqnarray}
where a prime indicates differentiation with respect to $z$. For
convenience, we use units with $\hbar^2/2m_e=1$, where $m_e$ is
the electron mass. The coupling matrix reads
\begin{eqnarray}
\label{gl::DefA}
        A_{\mu\nu}(z)&\!=\!&\!
        \int_0^{2\pi}\!\!\!d\varphi
        \int_{0}^{R(z,\varphi)}\!\!\!\!\!\!r\,dr
        \;\chi_\mu^*(r,\varphi;z)\;
    \frac{\partial\chi_\nu(r,\varphi;z)}{\partial z}.\quad
\end{eqnarray}
In the following we want to examine a cylindrical wire and its
sole classically unstable deformation - an axisymmetric one. In
this case the  geometry can be characterized by the radius
function $R(z)$ and the coupling matrix can be written as
$A_{\mu\nu}(z)= {\mathcal{A}}_{\mu\nu}\;\cdot R'(z)/R(z)$, where
${\mathcal{A}}_{\mu\nu}$ is now independent of $z$.
Eq.~(\ref{gl::gekoppeltesDGLsys}) simplifies to read
\begin{eqnarray}
\label{gl::gekoppeltesDGLsysR}
    &&\!\!\!\!\!\!\!\!
     C_\mu''+\left[E-E_\mu(R)\right]\,C_\mu\;=\;
        \sum_\nu \left[2\,\frac{R'}{R}{\mathcal{A}}_{\mu\nu}\,C_\nu'\,+
    \right.\nonumber\\
    &&\qquad\left.
        \left\{\!\frac{R''}{R}{\mathcal{A}}_{\mu\nu}-
        \left(\frac{R'}{R}\right)^{\!\!2}\!\!
\left({\mathcal{A}}_{\mu\nu}\!+\!
        {\mathcal{A}}^2_{\mu\nu}\right)\!\right\}C_\nu\right].\quad
\end{eqnarray}
In the adiabatic approximation, which is justified if the radius
varies slowly with $z$, the right hand side of
Eq.~(\ref{gl::gekoppeltesDGLsysR}) is neglected and the functions
$C_\mu(z)$ can be determined from a WKB approximation
\cite{Sta97a}. This procedure is not valid for small wavelength
modulations of the wire geometry that will be shown to limit the
maximal length. Hence, we have to solve
Eq.~(\ref{gl::gekoppeltesDGLsysR}) systematically.

%========================================================
%
%\section{Stability analysis}
%
%========================================================

To determine the stability of a cylindrical wire with given
radius $R_0$ and length $L$, we study small perturbations of its
geometry given by
\begin{eqnarray}
\label{RvonEps}
        R(z)&=&R_0\left(1+\varepsilon\;\sum_{q}\;b_q\;e^{i q z}\right).
\end{eqnarray}
The Fourier coefficients $b_q$ are chosen such that the total
volume of the wire is unchanged by the deformation. Other
physically reasonable constraints \cite{Sta99} are possible, but
lead to similar results. We calculate the scattering matrix of the
deformed nanowire connected to two cylindrical leads by matching
solutions of the Schr\"odinger equation
(\ref{gl::gekoppeltesDGLsysR}) at the boundaries. The density of
states (\ref{gl::DausS}) and the grand canonical potential
(\ref{gl::OmegaVonD}) are then derived as an expansion in the
deformation parameter $\varepsilon$.

%====================================================================
%
%                    R E S U L T S
%
%====================================================================

\begin{figure}[t]
    \vspace{0.5cm}
    \begin{center}
            \includegraphics[width=\columnwidth,draft=false]{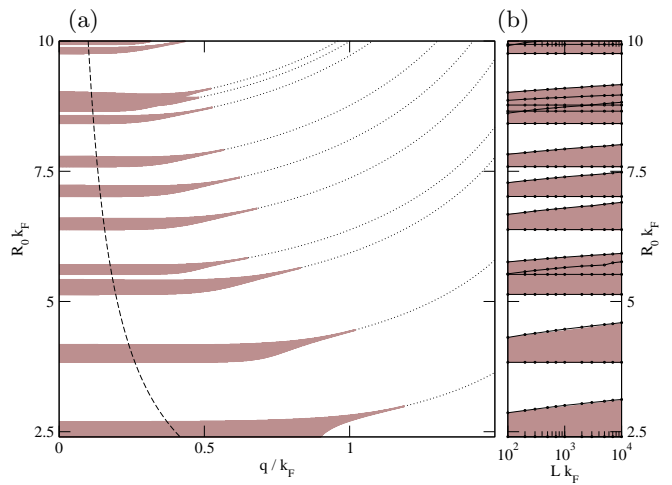}
    \end{center}
    \setlength{\unitlength}{1mm}
    \begin{picture}(0,0)
        \put(-35,68){(a)}
        \put(22,68){(b)}
    \end{picture}
    \vspace{-0.5cm}
    \caption[]{\label{abb::Stabdia}  a) Stability diagram
        for a cylindrical wire of length $Lk_F=1000$ at zero
        temperature. Shaded areas are unstable.
         b) unstable radii as a function of the wire length}
\end{figure}

The grand canonical potential $\Omega$ is found to read
\begin{eqnarray}
\label{gl::OmegaEntwicklung}
    &&\!\!\!\!\!\!
    \Omega(R(z))=\Omega^{(0)}(R_0)+\varepsilon^2\;\Omega^{(2)}(R_0)+{\mathcal{O}}(\varepsilon^3).\qquad
\end{eqnarray}
The leading order change $\Omega^{(2)}$ due to the deformation is
of second order in the Fourier coefficients $b_q$ and can be
written as
\begin{eqnarray}
\label{gl::ZerlegungOmega}
    &&\!\!\!\!\!\!
    \frac{\Omega^{(2)}}{L}=
    \sum_q|b_q|^2\alpha(q,R_0,L,T)
    +{\mathcal{O}}\left(\frac{1}{L}\right).
\end{eqnarray}
where the terms of order $1/L$ include nondiagonal contributions
which can be neglected if the wire is long enough.
Hence the function $\alpha(q,R_0,L,T)$ essentially determines the
stability of the nanowire and is therefore called \emph{stability
coefficient}. If $\alpha$ is positive for any value of the
perturbation wave vector $q$, all small deformations of the
cylindrical geometry lead to an increase of $\Omega$ and are
therefore suppressed.

First we want to discuss the limit of zero temperature.
Fig.~\ref{abb::Stabdia}(a) shows the stability diagram for a wire
of length $Lk_F=1000$ as a function of radius $R_0$ and
perturbation wave vector $q$ at $T=0$, where $k_F$ is the Fermi
wave vector. Shaded areas indicate a negative value of $\alpha$
and therefore instability. For comparison, the dashed line shows
the stability criterion for the Rayleigh instability ($q R_0=1$),
and the dotted lines show the criterion of the Peierls instability
\begin{eqnarray}
\label{gl::PeierlsBed}
    q &=& 2 k_F^{(\nu)}\;=\;2\sqrt{E_F-E_\nu(R_0)},
\end{eqnarray}
where $k_F^{(\nu)}$ is the Fermi wave vector of band $\nu$.

A sequence of stable and unstable radii can be extracted from
such a stability diagram. This reflects the existence of so
called magic stable radii, like they have been observed in
experiments. Fig.~\ref{abb::Stabdia}(b) shows the extent of the
regions of instability as a function of wire length, discussed
below.

\begin{figure}[t]
    \begin{center}
            \includegraphics[width=\columnwidth,draft=false]{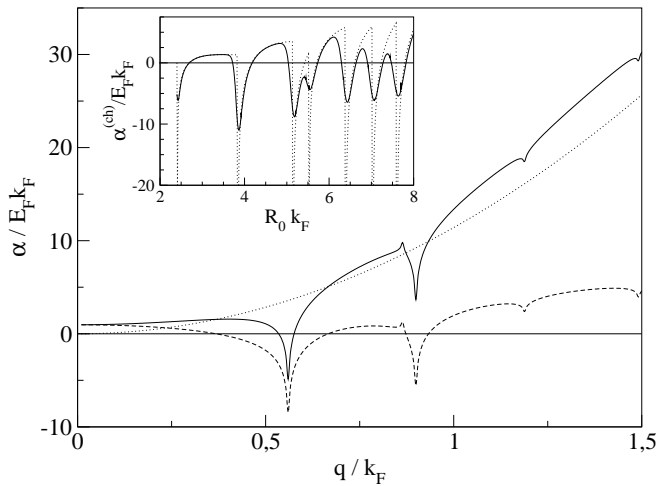}
    \end{center}
    \vspace{-0.5cm}
    \caption[]{\label{abb::Alpha} The stability coefficients $\alpha$ (solid line),
    $\alpha^{({\rm ch})}$ (dashed line) and $\alpha^{({\rm surf})}$ (dotted line) as a function of $q$ for
    $R_0 k_F=5.75$ and $Lk_F=1000$. Inset: $\alpha^{({\rm ch})}$ as a function of radius at
    $q =0.1k_F$ and temperature $T = 0$ (dotted line) and $T=0.02 T_F$ (solid line), $T_F$ being the Fermi
        temperature.   }
\end{figure}

In order to understand this result it is useful to distinguish
two contributions to $\alpha$ (Fig.~\ref{abb::Alpha}):
\begin{eqnarray}
\label{gl::ZerlegungAlpha}
    \alpha&=&\alpha^{\mbox{\tiny (ch)}}(q,R_0,L)
        + \alpha^{\mbox{\tiny (surf)}}(q,R_0).
\end{eqnarray}
The first term describes quantum effects due to the channel
structure of electronic transport and the second one includes
effects of surface tension and curvature energy.

For a given wire radius, $\alpha^{\mbox{\tiny (ch)}}$ as a
function of $q$ shows a series of distinct negative minima. With
increasing wire length, these minima get sharper and deeper,
resulting in a divergence for an infinitely long wire.
These instabilities arise from an effect similar to the Peierls
instability \cite{Pei56}. Nanowires are not purely one
dimensional, but due to quantization perpendicular to the wire
axis, electron transport is divided into distinct channels. Each
channel has a quadratic dispersion relation and starts to
contribute at a certain threshold energy, i.e., the eigenenergy
$E_\nu$ of the corresponding transverse mode. This results in a
sequence of quasi one dimensional systems with Fermi wave vectors
$k_F^{(\nu)}$, as defined in Eq.~(\ref{gl::PeierlsBed}). A
perturbation of the translational symmetry of the system results
in the opening of energy gaps in the electronic band structure.
If a gap opens at the Fermi surface, the total energy of the
system is lowered.

\begin{figure}[t]
    \begin{center}
            \includegraphics[width=0.8\columnwidth,clip,draft=false]{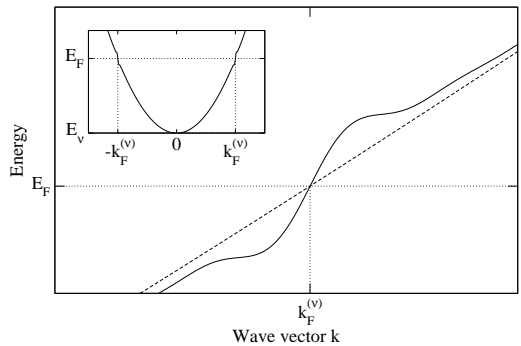}
   \end{center}
    \vspace{-0.5cm}
    \caption[]{\label{Abb::2} Dispersion relation of a
    channel $\nu$ (solid line)
    due to a perturbation of the wire with a wave vector $q=2k_F^{(\nu)}$. The inset shows the whole curve,
    while the main figure enlarges the vicinity of $k=k_F^{(\nu)}$.  The dashed line
    represents the unperturbed parabolic dispersion.  }
\end{figure}

The detailed calculation here is different from the standard
theory of the Peierls instability of a perturbed linear chain with
periodic boundary conditions \cite{Pei56}. There, the energy
change increases linearly with the length $L$ and $\alpha$
therefore is independent of $L$. Since the nanowire is part of a
much larger system including the leads, the longitudinal wave
vectors $k^{(\nu)}$ in the subbands are not restricted to
multiples of $\frac{2\pi}{L}$, like it is the case for an
isolated system. Instead, the entire system formed by the leads
and the wire determines the boundary conditions for the
$k^{(\nu)}$. Hence, an analogous Peierls problem arises if we
consider a linear chain of length $L_0$ that is perturbed by a
potential $\sim V\cos(2k_Fz)$ acting within an interval  of
length $L$ only, where $L\ll L_0$. In this case, the change of
the eigenenergies of states near $\pm k_F$ can be calculated by
perturbation theory and the resulting energy dispersion relation
is sketched in Fig.~\ref{Abb::2}. The dispersion relation is
smooth. Only in the limit $L \rightarrow \infty$ we recover the
well known Peierls results with a jump at $k_F^{(\nu)}$.

In the case of a single conducting channel, labeled $\nu$,
$\alpha^{\mbox{\tiny (ch)}}$ is given by
\begin{eqnarray}
    \label{eq::alphaCh::einKanal}
    \alpha^{\mbox{\tiny (ch)}}&\!\!=&\!\!\frac{16}{\pi}E_\nu k_F^{(\nu)}
    \!-\!\frac{4E_\nu^2}{\pi q}
    \int_{|2k_F^{(\nu)}\!-q|L}^{(2k_F^{(\nu)}\!+q)L}\frac{x-\sin(x)}{x^2}
    dx,
    \quad
\end{eqnarray}
and has a minimum for $q=2k_F^{(\nu)}$. For several open channels,
$\alpha^{\mbox{\tiny (ch)}}$ includes terms analogous to
Eq.~(\ref{eq::alphaCh::einKanal}) for each channel, as well as
similar but more complicated terms that take into account
coupling between any two channels having the same azimuthal
symmetry. This leads to additional minima for
$q=k_F^{(\mu)}+k_F^{(\nu)}$, that are less pronounced than those
for $q=2k_F^{(\nu)}$. Away from the Peierls-type instabilities,
$\alpha^{\mbox{\tiny (ch)}}$ is positive and the nanowire is
stabilized by electron shell effects.

The second term in Eq.~(\ref{gl::ZerlegungAlpha}) also tends to
stabilize the wire. It is given by
\begin{eqnarray}
\label{gl::AlphaSurf}
    \alpha^{\mbox{\tiny (surf)}}&\!\!=\!\!&-
    \frac{2}{\pi}\sum_\nu
        k_F^{(\nu)}\left[{(\mathcal{A}}^2)_{\nu\nu}+\sum_\mu|{\mathcal{A}}_{\mu\nu}|^2\right]q^2,
\end{eqnarray}
with sums running over all open channels. A proportionality to
$q^2$ is expected for the change of the surface and curvature
energy of the wire due to a deformation with wave vector $q>0$.
Using the explicit expressions for the coupling matrix elements
one can show that $\alpha^{\mbox{\tiny (surf)}}(q)$ is always
positive. Note that the negative surface contribution at $q=0$
that leads to the well-known Rayleigh instability is contained in
$\alpha^{({\rm ch})}$.

Now a coherent description of the effects determining the
stability of a nanowire arises. Surface tension and curvature
energy tend to stabilize the nanowire for deformations of small
wavelength since the cost in energy increases $\propto q^2$. On
the other hand, the classically expected Rayleigh instability for
long wavelength perturbations is suppressed by electron shell
effects that depend strongly on the wire radius. The channel
structure of the electronic transport results in several
Peierls-type instabilities for perturbations of wave vector
$q=2k_F^{(\nu)}$ that become more pronounced with increasing wire
length. The competition of these instabilities with the
stabilizing effects of surface tension and curvature energy
limits the maximal length of stable nanowires
(Fig.~\ref{abb::Stabdia}). Nevertheless, stable wire geometries
persist up to lengths greater than $Lk_F = 10^4$ (i.e. $L\sim
1\mu m$ for Na and Au) which is encouraging for the technological
prospects of nanowires.

%============================================
%
%\section{ Finite temperature }
%
%============================================

\begin{figure}[t]
    \vspace{0.5cm}
    \begin{center}
            \includegraphics[width=\columnwidth,draft=false]{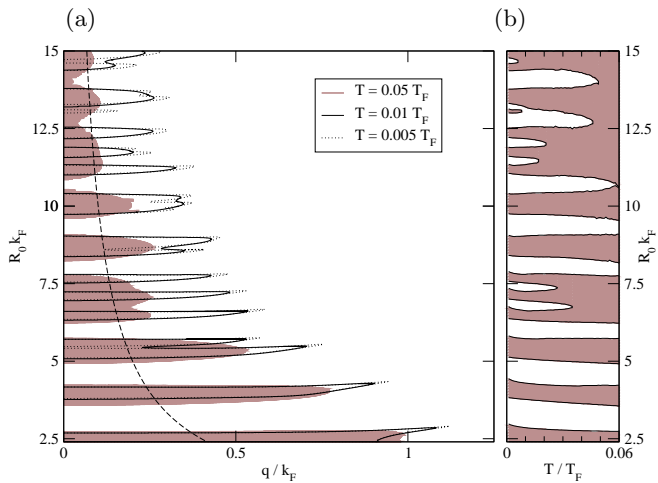}
    \end{center}
    \setlength{\unitlength}{1mm}
    \begin{picture}(0,0)
        \put(-35,68){(a)}
        \put(22,68){(b)}
    \end{picture}
    \vspace{-0.8cm}
    \caption[]{\label{Abb::StabdiaT} (a) Stability diagram
        for a cylindrical wire of length $Lk_F=1000$ at three different
        temperatures. The shaded area indicates instability at
        $T=0.05 T_F$, the solid and the dotted lines show the
        contours of the unstable regions for $T=0.01T_F$ and
        $T=0.005T_F$ respectively, $T_F$ being the Fermi
        temperature.
        (b) unstable radii as a function of temperature}
\end{figure}

Now we turn to the effect of temperature. Increasing temperature
results in a broadening and a flattening of the minima describing
the Peierls instabilities (Inset Fig.~\ref{abb::Alpha}). The
stability diagram of a cylindrical wire of length $Lk_F=1000$ is
shown in Fig.~\ref{Abb::StabdiaT}(a) for three different
temperatures. The Peierls instabilities disappear at higher
temperature and the stability diagram approaches that of the
classical Rayleigh instability, which is shown by the dashed line.
Nevertheless islands of stability persist up to very high
temperatures. Fig.~\ref{Abb::StabdiaT}(b) shows the extent of the
unstable regions as a function of temperature and the increase of
unstable configurations with growing temperature can be seen. The
variation in the maximum temperature up to which the wires remain
stable reflects the supershell structure observed in
Ref.~\cite{Yan00}.

In our calculations the free electrons are kept in the wire by an
infinitely high potential well ("hard-wall"). Since the electrons
in a real metal are not so strongly confined, a more physical
boundary condition would be a potential well of finite height
depending on the electron work function. The main influence of
this "soft-wall" boundary condition is to increase the effective
radius of the cylindrical wire relative to the "hard-wall" case
\cite{Gar96}. This has to be kept in mind when comparing our
predictions to experiments.

Recently experiments found evidence of nanowires made out of
dimerized chains of gold atoms \cite{Cso03}. This is not in
contrast to our results, as in this experiment surrounding
molecular hydrogen had a stabilizing effect on the wire. Also the
Jellium model used for our calculations is at its limit of
validity for a chain of atoms.

%===============================================================
%
%    C O N C L U S I O N S
%
%===============================================================

In conclusion, we have presented a fully quantum mechanical
stability analysis of cylindrical nanowires, using the Jellium
model. The stability of a nanowire is determined by the interplay
of surface tension, shell effects and the Peierls instability.
Our calculated stability diagrams prove the existence of certain
stable magic radii like they have been found in experiments and
give an explanation for the observed cylindrical geometry of
these nanowires.

We thank C.~A.~Stafford, F.~Kassubek and J.~B\"urki for valuable
discussions. This research has been supported by the DFG through
SFB 276 and the EU Network DIENOW.

\bibliography{refs}

\end{document}